\documentclass{emulateapj}

\shorttitle{An unusual Quasar SED}
\shortauthors{Th. Boller, P. Schady and T. Heftrich}

\begin{document}

\title{XMM-Newton, Swift and ROSAT observations of LBQS 0102-2713}

\author{Th. Boller\altaffilmark{1}, P. Schady\altaffilmark{1} and T. Heftrich\altaffilmark{2}}
\affil{$^1$Max-Planck-Institut f\"ur extraterrestrische Physik Garching, PSF 1312, 85741 Garching, Germany\\
       $^2$GSI Helmholtzzentrum f\"ur Schwerionenforschung GmbH, Planckstr. 1, 64291 Darmstadt, Germany
       }
\email{bol@mpe.mpg.de}

\begin{abstract}
\noindent
We have analyzed the first XMM-Newton, Swift and archival ROSAT PSPC observations of the quasar LBQS 0102-2713.
The object was selected from the ROSAT archive as being notable due to the steep soft X-ray
photon index and due to the UV brightness based on HST and optical spectroscopic observations.
The first XMM-Newton observations carried out in December 2009 and the first Swift observations
from 2010 have confirmed the steepness of the soft X-ray photon index, which  ranges between 3.35 and 4.41 for the
different XMM-Newton and ROSAT detectors, the UV brightness of the source  and the absence of 
significant absorption by neutral hydrogen.
The new data allow a combined spectral fitting to the Swift UVOT and the XMM-Newton/ROSAT data which results
in a huge luminosity of  $\rm  (6.2\pm0.2)\times10^{47}\ erg\ s^{-1}$ and $\rm \alpha_{ox}$ values
ranging between $\rm  (-1.87\pm0.11)\ and\ (-2.11\pm0.12)$.
The nature of the soft X-ray emission can be explained as local comptonized emission of the UV disc photons in
the pseudo-Newtonian potential.
The black hole mass is estimated from the
Mg II line and translates into an Eddington ratio of  $\rm   L/L_{edd} = 18^{+33}_{-12}$.
For the dimensionless electron temperature of the plasma cloud $\rm \theta = kT_e/ m_{e} c^2$ we derive an upper limit of
about 10 keV.
\end{abstract}

\section{Introduction}
\noindent
LBQS 0102--2713 is a Narrow-Line Quasar at a redshift of $\rm  z=0.780\pm 0.005$ 
with a B magnitude of $\rm 17.52\pm0.15$ (Hewett et al. 1995).
The object was selected from the catalogue of ROSAT pointed observations as being notable and 
important due to its steep photon index for a quasar (Boller et al. 2009).
The photon index obtained from a power law fit was
$\rm \Gamma=(6.0\pm1.3)$ for an $\rm N_H$ value of $\rm (4.8\pm1.5)\times10^{20}\ cm^{-2}$.
The photon index  remained steep at about 3.5 when the $\rm N_H$ value was fixed to the Galactic value
of $\rm 1.2\times10^{20}\ cm^{-2}$  (Dickey and Lockman, 1990).
The $\rm \alpha_{ox}$  was steep with a value of about -2.3, comparable to the steepest values detected
in BAL quasars. However, the maximum intrinsic absorption 
was at least a factor of about 20 lower compared to BAL quasars. 

\noindent
The Mg II line at 2800 \AA\ rest frame has a FWHM of about 2200 $\rm
km\ s^{-1}$ (c.f. Fig. 4 of Boller et al. 2009). 
This value is very close to the somewhat arbitrary line between Narrow-Line Seyfert 1 Galaxies
(NLS1s) and broad line Seyfert galaxies of 2000 $\rm km\ s^{-1}$ following
the definition of Osterbrock \& Pogge (1985). As the $\rm 2000\ km\ s^{-1}$ 
arbitrary line is luminosity dependent (Shemmer et al. 2008), LBQS 0102-2713 
can be considered as a luminous NLS1.
In addition there is strong UV Fe II multiplet emission between about
2200 and 2500 \AA\ in the rest frame. All this is typical for NLS1s 
(c.f.  Tr\"umper \& Hasinger 2008).

\noindent
There are no significant indications that the
object is intrinsically X-ray weak, in contrast to the argument used for
PHL 1811 by Leighly et al. (2007). 
In the case of an intrinsically
weak X-ray source one expects weak low ionization or semiforbidden
UV lines. The rest-frame EW values of LBQS 0102-2713 for the blend of
$\rm Ly\beta$ and the O~VI lines, and the $\rm Ly\alpha$ and the N~V lines are
about 12 and 50 \AA, respectively when comparing these values with quasar composites
the source appears not to be intrinsically X-ray weak. 
For the $\rm Ly\beta$ plus O~VI lines Brotherton et al. (2001) give an EW value of 11 \AA\ in the
rest frame and for the blend of $\rm Ly\alpha$ and N V a value of 87 \AA.
Similar values can be found in the quasar composites of Vanden Berk et al. (2001)
and Zheng et al. (1997). 

\noindent
In this paper we analyse the first XMM-Newton and Swift observation of the source
and give tighter constrains on the important $\rm \alpha_{ox}$ and soft X-ray photon index
values, the black hole mass, UV-soft X-ray luminosity and the Eddington ratio.

\section{X-ray, UV observations and data analysis}
\noindent
LBQS 0102-2713 was observed for the first time with XMM-Newton during revolution 1829 
in December 2009 for 22218 seconds. 
The Swift XRT and UVOT observation were taken  in August 2010 
with total exposure times of 5079 seconds for the UVOT  and 
5151 seconds for the XRT measurements.
ROSAT PSPC observations are available for three pointings in January, June and
December 1992 with 6157, 2191, and 6724 second exposures, respectively.
In Table 1 we list the count rates and expsosure times for the individual detectors.
The unobscured flux and luminosity is derived in the Section 4.

\noindent
The XMM-Newton observations were processed  using SAS 10.0.0 (xmmsas\_20100423\_1801\-10.0.0).
The spectra and response values have been calculated by using the latest pn and MOS chains. 
We have fitted the XMM-Newton spectra in the energy range between 0.3 and 2.0 keV where the source
is not background dominated. We note that we have carefully ignored high background flaring events and find that
more than 90 per cent of MOS observations and more than 65 per cent of the pn observations
could be used in the subsequent data analysis.
The spectra are binned using the {\it grppha} command. Each bin contains at least 30 counts. 
The Swift data were processed via the standard data analysis procedures.
The data analysis of the archival ROSAT observation are described in detail in Boller et al. (2009)
and are based on the {\it xselect} command interface. 
The spectral fitting results are obtained using {\it XSPEC, version 12.4.0} (Arnaud et al. 1996).

\begin{table}
\begin{center}
\caption{Count rates and exposure times for the XMM-Newton EPIC pn, MOS1, MOS2 and the
Swift UVOT detectors.}
\begin{tabular}{lll}
\tableline\tableline
Instrument      &   count rate                      &        exposure time \\
                &   [$\rm counts\ s^{-1}$]	    &	     [ks]          \\
pn              &   $\rm (1.20\pm0.01)\times 10^{-1}$	  &  15.150         \\
M1              &   $\rm (2.26\pm0.13)\times 10^{-2}$	  &  20.260         \\
M2              &   $\rm (2.10\pm0.12)\times 10^{-2}$	  &  20.780         \\
ROSAT           &   $\rm (4.80\pm0.29)\times 10^{-2}$	  &  12.540         \\
Swift v         &   $\rm  1.70\pm0.06               $	  &  1.073          \\
Swift u         &   $\rm  5.30\pm0.15               $	  &  1.017         \\
Swift uvw1      &   $\rm  3.10\pm0.08               $	  &  1.218         \\
Swift uvw2      &   $\rm  3.00\pm0.08               $	  &  1.245         \\
\tableline
\end{tabular}
\end{center}
\end{table}

\section{XMM-Newton, ROSAT and SWIFT XRT X-ray data}

\subsection{LBQS 0102-2713 as a steep soft X-ray spectrum quasar}
\noindent 
The ROSAT and XMM-Newton spectra have been fitted simultaneously, but allowing the
respective values for the photon indices and normalizations to vary, given the large time difference
between the observations.
Spectral fitting was performed up
to 2 keV, since above that energy the emission is background dominated.  A simple power law 
model with neutral absorption results in a statistically acceptable fit. An $\rm N_H$
value is obtained from the ROSAT data of $\rm 2.85\times10^{20}\ cm^{-2}$, 
with
$\rm  \Delta \chi^{2}=2.71$ confidence levels ranging between 
$\rm  (2.2\ and\ 3.6)\times10^{20}\ cm^{-2}$, indicating that absorption above the Galactic value is required.
In the simultaneous fit
to the XMM-Newton and ROSAT data the $\rm N_H$ value was fixed to the value obtained from
ROSAT observations.
At the $\rm  \Delta \chi^{2}=2.71$ confidence level the photon indices range
between 3.35 and 4.41.
The Swift XRT data exhibit the lowest count rate statistics with
$\rm (4.0\pm0.9)\ 10^{-3}\ counts\ s^{-1}$ and are not included into the joint fit
as the power law and normalization values remain unconstrained.
The normalization values obtained from the XMM-Newton and ROSAT data indicate no 
significant variability between the ROSAT observations obtained in 1990 and the 
XMM-Newton data from 2009  (c.f. Table 2).
Similar to the ROSAT PSPC light curves (c.f. fig. 1 of Boller et al. 2009)
no significant short timescale variability is detected within the XMM-Newton
observations.

\begin{table}
\begin{center}
\caption{
$\rm  \Delta \chi^{2}=2.71$ confidence levels
of the power law fit parameters obtained 
from XMM-Newton and ROSAT. The ROSAT and XMM-Newton spectra have been fitted
simultaneously, but allowing the parameters to vary, given the large time difference
between the observations. The $\rm N_H$ value is fixed to the value obtained from
the ROSAT data due to the lower energy coverage of ROSAT compared to XMM-Newton.
The normalization is given in units of $\rm photons\ cm^{-2}\ s^{-1}\ keV^{-1}$ at 1 keV.}
\begin{tabular}{llll}
\tableline\tableline
Instrument      &   Photon index  &   norm	 &    	$\rm N_H$        \\
                &		  & [$\rm 10^{-5}]$&  [$\rm 10^{20}\ cm^{-2}$] \\
pn              &   3.86-4.20	  &   2.59-3.37  &     2.85          \\
M1              &   3.49-4.02	  &   3.12-4.42  &    		         \\
M2              &   3.35-3.85	  &   2.90-4.08  &    		         \\
ROSAT           &   3.92-4.41	  &   1.08-3.28  &    		         \\
\tableline
\end{tabular}
\end{center}
\end{table}

\noindent
In Fig. 1 we show the power law fit to the XMM-Newton and ROSAT data. 
The range of photon indices listed in Table 2 belong to the steepest values 
obtained from quasar X-ray spectra.
For comparison we list below the mean photon index plus the corresponding errors
for samples of other quasars reported in the literature.
For the sample of luminous SDSS quasars of Just et al. (2007) the corresponding value is
$\rm  1.92^{+0.09}_{-0.08}$.
The values obtained from Chandra observations of SDSS quasars by Green et al. (2009)
and on high redshift quasars by Vignali et al. (2003) are  
$\rm  1.94\pm0.02$, and $\rm  1.84^{+0.31}_{-0.30}$, respectively.
The mean photon index and the related errors for the highest redshift SDSS quasars
obtained by Shemmer et al. (2006) is $\rm  1.95^{+0.30}_{-0.26}$.

\noindent
To our knowledge there are only a few other quasars with photon indices as steep as LBQS 0102-2713.
Grupe et al. (1995) report on an extreme steep power law slope of about 7 in WPV007.
The object is also of general interest due to the extreme X-ray variability which
is most likely due to changes in the BAL flow as a result of absorption (e.g. 
Leighly et al. 2006).

\noindent
Another example is the photon index of 4.2 derived by 
Komossa et al. (2000)  
for the NLS1 galaxy  RX J0134.3-4258.
Molthagen (1998) reports on a
photon index of about 4.3 in  RXJ0947.0+4721, and 
George et al. (2000) obtained a photon index of
$\rm 4.18^{0.82}_{1.1}$ in PG 0003+199.

\begin{figure}
\includegraphics[angle=-90,scale=0.35,clip=]{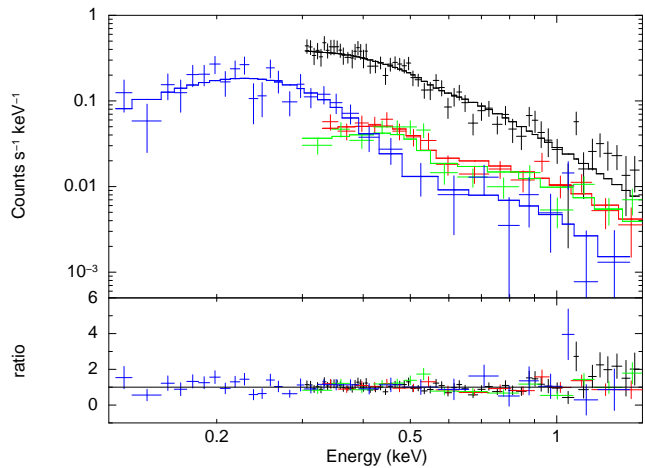}
\caption{
Joint fit to the ROSAT and  XMM-Newton EPIC pn, MOS1, MOS2 spectra. A simple power law
model with neutral absorption has been applied. All parameters are free in the fit, except
the neutral absorption, which was obtained from the ROSAT observations and which gives the most
precise value due to its low energy coverage. 
The color coding is as follows: 
ROSAT PSPC (blue), 
EPIC pn (black),
MOS1  (red), and
MOS2 (green).
}
\end{figure}

\subsection{$\alpha_{ox}$ determination}
\noindent
The 2 keV rest-frame flux density is determined following Hogg (2000) and
Weedmans's Quasar Astronomy (1986). 
For a photon index not equal to 2 the rest frame 2 keV flux density is
given by
$\rm f_{2keV} = f(0.5-2.0)\times ((1+\alpha_x) / (1+z)^{\alpha_x}))\times
((\nu_{2keV}^{\alpha_x} / (\nu_{2keV}^{\alpha_x + 1} - \nu_{0.5keV}^{\alpha_x +
1}))$.  
The  unabsorbed flux in the 0.5-2.0 keV energy band is
$\rm  (1.04\pm0.06)\times10^{-13}$ for MOS1,
$\rm  (9.21\pm0.53)\times10^{-14}$ for MOS2,
$\rm  (8.82\pm0.07)\times10^{-14}$ for pn, and
$\rm  (6.73\pm0.41)\times10^{-14}$
$\rm  erg\ cm^{-2}\ s^{-1}$ for ROSAT.
Using the full range of unabsorbed 0.5-2.0 keV flux measurements and the photon indices listed
in Table 2, one
obtains a  2 keV flux density that ranges between $\rm  (8.83\pm0.52)\times10^{-32}$ and
$\rm  (2.05\pm0.12)\times10^{-31}\ erg\ cm^{-2}\ s^{-1}\ Hz^{-1}$.
From the Swift UVOT data we derive an extinction corrected
flux at 2500 \AA\  rest frame wavelength of 
$\rm (6.35\pm0.02)\times10^{-27}\ erg\ cm^{-2}\ s^{-1}\ Hz^{-1}$.
The corresponding  $\rm  \alpha_{ox}$ values range between
$\rm  (-1.87\pm0.11)\ and\ (-2.11\pm0.12)$, respectively.
These values are somewhat lower compared to the -2.2 value obtained
from the ROSAT data in Boller et al. (2009) due to the steeper photon index assumed of 6. 
The range in the $\rm \alpha_{ox}$ values obtained from the additional first XMM-Newton and Swift
observations provide more precise measurements. 

\noindent
Just et al. (2007) argued that $\rm  \alpha_{ox}$ decreases with increasing 2500\ \AA\ luminosity 
density. In their fig. 7
the $\rm  \alpha_{ox}$ values range between about -1.0 and -2.2, and 
the $\rm  log\ L_{2500 \\A\ } = l_{UV}$ 
luminosity density ranges between about 27.8 and 32.5. LBQS 0102-2713 exhibits a luminosity 
density at 2500\ \AA\  of $\rm  L_{2500 \\A\ }  = (5.67\pm0.02)\times10^{31}$
$\rm  erg\ s^{-1}\ Hz^{-1}$
corresponding to a value of   $\rm  l_{UV}=(31.75\pm0.001)$.
Therefore the source is located at the higher end of the $\rm  l_{UV}$ distribution.
The $\rm  \alpha_{ox}$ values given above on the other hand 
are at the lower end compared Just et al. (2007).
The corresponding $\rm   \Delta\alpha_{ox}$ 
values are ranging between  -0.05 and -0.5. All this is
further supporting the trend whereby $\rm  \alpha_{ox}$ decreases with increasing UV luminosity density.

\section{SED spectral fitting}

\noindent
We have fitted the Swift UVOT, XMM-Newton and ROSAT data with
an accretion disc spectrum {\it diskpn} and the {\it compTT} model including 
absorption by neutral hydrogen
{\it TBabs} and the corresponding extinction value {\it redden}. 
As absorption above the Galactic value is required, the $\rm  N_H$ value is allowed
to vary to obtain the best value for the total luminosity.
The
$\rm  \Delta \chi^{2}=2.71$ confidence levels range between $\rm  (2.3\ and\ 3.9)\times
10^{20}\ cm^{-2}$, slightly above the values derived from the ROSAT power law fit.
The X-ray absorption by the ISM is accounted for by the {\it  XSPEC} model {\it  TBabs} 
(Wilms et al. 2000).
The {\it  redden} model describes  the infrared, optical, and
ultraviolet extinction  (Cardelli et al. 1989).
The {\it  diskpn} model available within {\it  XSPEC} is an extension of the {\it  diskbb} 
model of 
Mitsuda et al. (1984) and Makishima et al. (1986) and 
 describes the black body emission of a disk in the pseudo-Newtonian potential 
(for details see Gierlinski et al. 1999). 
A part of the thermal disc photons are Comptonized in a hot plasma described by the {\it  compTT} model
Titarchuk (1994). Both models  components are available within the version of XSPEC that we used.
The phenomenological model gives an acceptable fit to the Swift UVOT, XMM-Newton and ROSAT data. 
The temperature of the {\it  diskpn} model is $\rm  30\pm4\ eV$. The inner radius is fixed
to $\rm  10\ R_G$ and the normalization is $\rm  171\pm5$.
The normalization is given in units of $\rm  (M^2 cos\ i)/(D^2 \beta^4)$.
The input soft photon temperature of the {\it  compTT} model is fixed to the temperature of the 
{\it  diskpn} spectrum and the normalization is $\rm  (1.80\pm0.21)\times10^{-4}$. 
As the high energy cut-off is not detected in the data, the electron temperature has been fixed 
to 50 keV. The plasma optical depth is $\rm  (4.20\pm0.69)\times10^{-2}$ which is a lower limit given
the fact the the high energy cut-off is not detected.
The reduced $\rm \chi^2$ value is 1.09 for 114 d.o.f.

\begin{figure}
\includegraphics[angle=-90,scale=0.33,clip=]{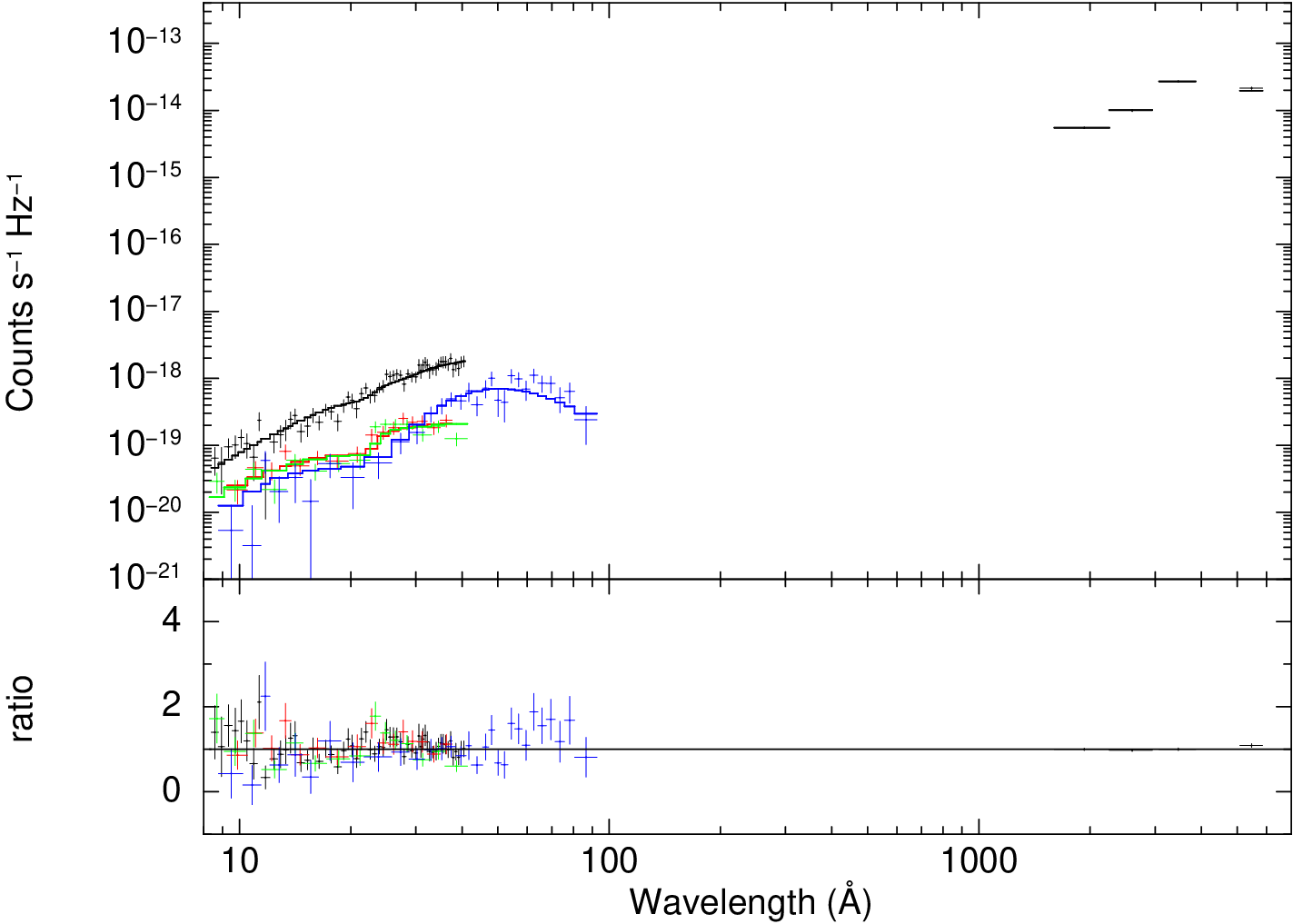}
\includegraphics[angle=-90,scale=0.33,clip=]{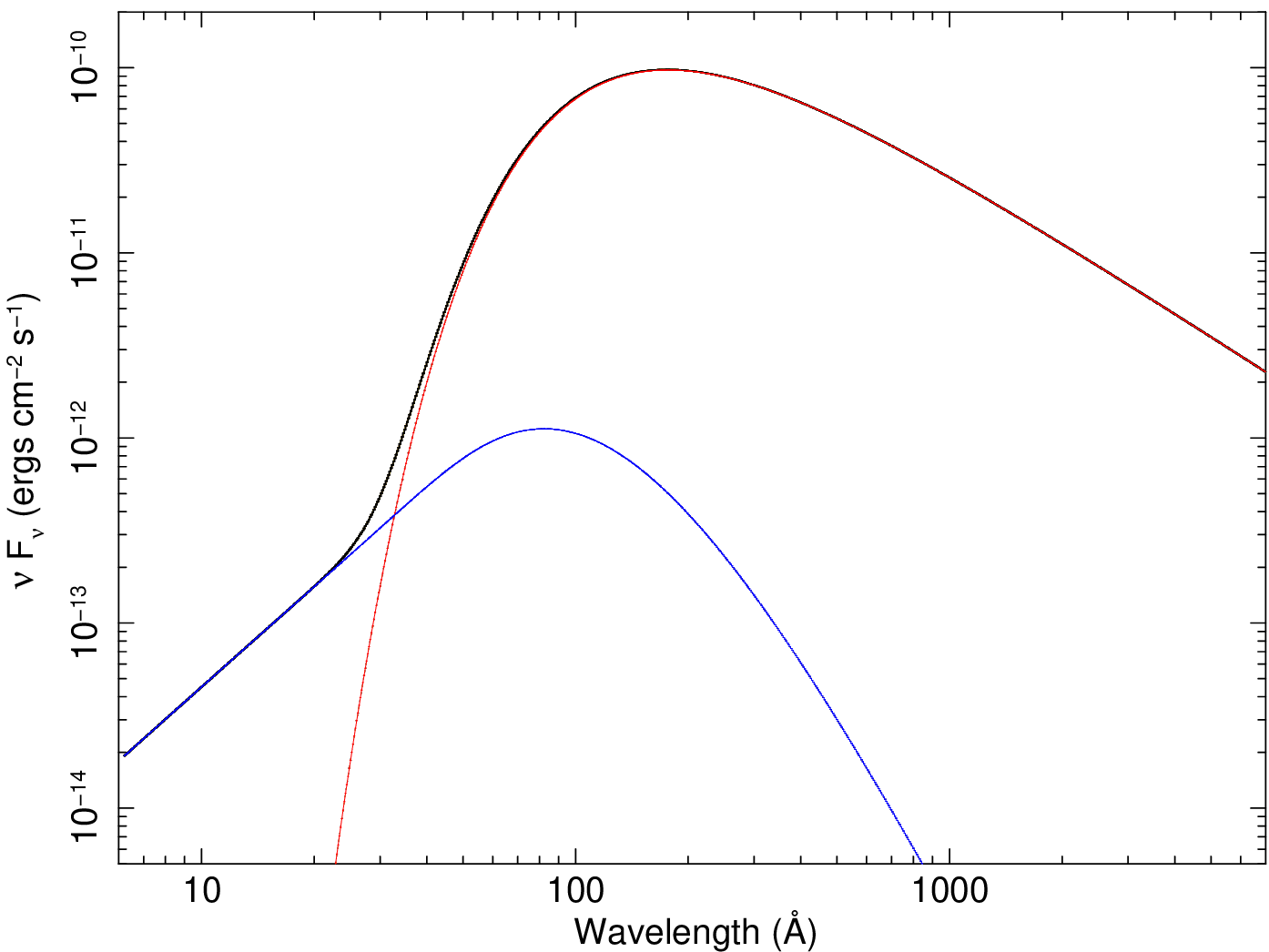}
\caption{
Upper panel:
Fit to the Swift UVOT, XMM-Newton and ROSAT data for an accretion disc spectrum which is
partially Comptonized. The UVOT data describe the emission from the accretion disc
and the soft X-ray flux is arising due to Comptonization of parts of the accretion disc by a hot electron
layer. 
Lower panel:
The unfolded model spectrum. Note the strong UV luminosity and the X-ray luminosity
close to 2 keV, which differ by about 4 orders of magnitude in the $\rm \nu\ f_{\nu}$ parameter space.
} 
\end{figure}

\noindent
The unabsorbed luminosity derived from the spectral fit is $\rm  6.20\times 10^{47}\ erg\ s^{-1}$
in the energy range between 0.001 and 2 keV. 
Using the two extreme normalization values for each relevant model component we derive a 
lower value and upper value for the unabsorbed luminosity of $\rm  6.01\times 10^{47}\ erg\ s^{-1}$
and $\rm 6.42\times 10^{47}\ erg\ s^{-1}$.  
We note that the luminosity is comparable to the mean value of the most luminous quasars obtained by  Richards
et al. (2006). 
Assuming an Eddington ratio of 1, this translates to a lower limit of the black hole mass of
 $\rm  (4.50^{+0.15}_{-0.17})\times10^{9}\ M_{\sun},$ 
which would appear on the upper mass scale for black holes.
However, as it is known for the lower luminosity NLS1 analogues, the Eddington ratio might exceed 1 for
objects with steep soft X-ray photon indices and narrow optical line widths.

\noindent
A more reliable mass  and Eddington ratio estimation can be obtained using the Mg~II 
rest frame line width
which is 2200 $\rm km\ s^{-1}$ (Boller et al. 2009) and the optical observation
obtained by Morris et al. (1991). 
Following equ. 10 of Wang et al. (2009), the Mg II FWHM line
width and the $\rm  \lambda\ L_{\lambda}\ 3000\ \AA$ value of $\rm  3.2\times10^{45}\ erg\ s^{-1}$ 
translate into a black hole mass of $\rm  (2.51^{+4.34}_{-1.60})\times10^8\ M_{\sun}$ 
assuming a luminosity distance of $\rm  D_L=1.5\times10^{28}\ cm$. 
The corresponding Eddington luminosity is 
$\rm  (3.46^{+5.99}_{-2.20})\times10^{46}\ erg\ s^{-1}$.
Taking into account the lower and upper values for the unabsorbed luminosity 
and the Eddington luminosity, respectively, the Eddington ratio is
$\rm  L/L_{edd} = 17.92^{+33.0}_{-11.5}$. 
Eddington ratios have been estimated by several authors. Onken and Kollmeier (2008) found Eddington ratios
for SDSS quasars ranging between 0.01 and 1. 
Shemmer et al. (2004) derive values for the Eddington ratio ranging between 0.14 and 1.71 (their Table 2).
We note that LBQS 0102-2713 appears to exhibit one of the highest Eddington ratios 
 measured 
from accreting black holes so far.

\section{Summary}
\noindent
The NLS1 quasar LBQS 0102-2713 exhibits an unusual parameter combination derived from archival ROSAT and the
first XMM-Newton and Swift observations that can be summarized as follows:
(i)    the soft X-ray photon index ranges between 3.35 and 4.41 and belongs to the steepest values observed so far;
(ii)   the $\rm \alpha_{ox}$ value ranges between
       $\rm  (-1.87\pm0.11)\ and\ (-2.21\pm0.12)$,  similar to the highest values observed in BAL quasars
       (c.f. table 3 of Gallagher et al. 2006);
(iii)  in contrast to BAL quasars, LBQS 0102-2713 is not significantly absorbed, the most precise measurement,
       which comes from 
       ROSAT observations, is $\rm 2.85\times10^{20}\ cm^{-2}$, at least two
       orders of magnitude lower compared to BAL quasars;
(iv)   the 2 keV monochromatic luminosity is comparable to the mean value for quasar SED's, however the object is not
       intrinsically X-ray weak, in contrast to PHL 1811 which shows similar observational properties;
(v)    the UV-X-ray luminosity is $\rm  6.20^{+0.22}_{-0.19}\times10^{47}\ erg\ s^{-1}$, about two 
       orders of magnitude above the mean
       of quasar SED's, and the ratio of the UV peak to 2 keV X-ray luminosity is about  $\rm 10^4$ 
       in the $\rm \nu F_{\nu}$ space;
(vi)   the Eddington ratio appears extremely high with $\rm L/L_{edd} = 17.92^{+33.0}_{-11.5}$
       compared to other quasars studies.
(vii)  no X-ray emission is detected above 2 keV and the upper limit obtained from the first XMM-Newton observations
       is $\rm 2\times10^{-14}\ erg\ cm^{-2}\ s^{-1}$.

\section{Discussion}
\noindent
We have fitted the Swift UVOT, XMM-Newton and the ROSAT data with an accretion disc spectrum in the pseudo-Newton
potential. 
The temperature of the seed photons is $\rm  T_{seed,max} = (30\pm4)\ eV$.
A part of the thermal disc photons are comptonized in a plasma cloud with some Thomson depth $\rm \tau_{electrons}$ and
a temperature $\rm T_{electrons}$.
The mean change in photon frequency is given by
$\rm  \Delta \nu / \nu = (4kT_e - h\nu) / m_e\ c^2$ (Titarchuk and Hua 1997).
For a thermal distribution of electrons the dimensionless electron temperature is $\rm \theta = kT_e/ m_{e} c^2$. 
The scattered input and output photon energies are related by $\rm \epsilon_{out,N} = (1 + 4\theta)^N\epsilon_{in} =
 3\theta$, where N is the number of scattering events which depend on the plasma optical depth. 
The shape of the spectrum is determined by the Comptonization parameter $\rm  y = kT_e / m_e c^2 N$ 
(Titarchuk and Hua 1997). 
The relation between the photon index and the electron temperature is
$\rm \Gamma = (ln \tau) / ln(1+\theta)$. For a given optical depth a steep power law means that the electron temperature is
low. The comptonized UV seed photons give rise to the observed X-rays as detected with ROSAT and XMM-Newton.
With this model we are able for the first time to explain the UV and X-ray emission.
As no thermal energy cut-off is detected in the Comptonized spectrum, the plasma 
optical depth $\rm \tau_{electrons}$ and the 
electron temperature $\rm T_{electrons}$ cannot be reliably determined from the {\it compTT} model.
However the Comptonization parameter y can be derived from the power law spectrum. 
The $\rm  \Delta \chi^{2}=2.71$ confidence levels
for the photon indices listed in Table 1 translate to y values ranging between
$\rm  4.5\times10^{-2}$ and $\rm  7.9\times10^{-2}$.

\noindent
As the thermal cut-off is not detected, only some estimates on the temperature of the plasma can be made.
We note that the source is accreting above the Eddington limit and that in such cases outflowing winds are expected with some
significant optical depth. 
Following Rybicky and Lightman (1979) the relation between the Comptonization parameter, 
the dimensionless electron temperature
and the plasma optical depth is $\rm y = 4\theta \times \tau_{electrons}$. 
Assuming a plasma optical depth $ \ge$ 1, an upper limit for $\rm \theta$ can be derived.
Using the y values given above and $\rm  \tau_{electrons} = 1$  the upper-limit on $\rm  \theta$  
ranges between  $\rm  1.1\times 10^{-2}$ 
and $\rm  2.0\times 10^{-2}$, corresponding to electron temperatures ranging between 5.6 and 10.2 keV.
To check that this is consistent with our spectral data, we 
fixed the plasma optical depth to $\rm  \tau_{electrons} = 1$  in the spectral fit described in Section 4, and we
derived an electron temperature of $\rm  (6.8\pm0.4)\ keV$, in agreement with the analytical calculations. 
Longer X-ray observations are required to finally detect the thermal cut-off and to precisely determine the 
coronal parameters.


\vskip 0.2cm
\noindent
\begin{acknowledgements}
\noindent
The authors would like to thank the anonymous referee for her/his detailed and extremely helpful
report.
TB is grateful for critical reading of the paper by A. M\"uller and M. Gilfanov as well as
constructive discussions with S. Nayakshin.
TB especially thanks C. Done for fruitful discussions on the SED spectral fitting of the source and the 
importance of Comptonization to explain the X-ray emission in super-Eddington sources.
TB thanks S. Immler for the discussions on the importance of the Swift satellite for observing UV bright 
distant AGNs. 
\end{acknowledgements}

\vskip 0.2cm

\noindent
{\it Facilities:} \facility{ROSAT, HST FOS, XMM-Newton, Swift}.

\end{document}